\documentclass[12pt,aps,ams,amsfonts,nofootinbib]{revtex4}

\usepackage[dvips]{graphics}
\usepackage{graphicx}
\usepackage{amssymb}
\usepackage{subfigure}
\usepackage{stackrel}
\usepackage{enumitem}
\usepackage{amsmath}
\usepackage{amsfonts}
\usepackage{bm}
\usepackage{color}
\usepackage{verbatim} 
\usepackage{tcolorbox}
\usepackage[normalem]{ulem}
\usepackage{hyperref}
\usepackage{comment}

\newlist{subquestion}{enumerate}{1}
\setlist[subquestion,1]{label=(\alph*)}

\newcommand{\vect}[1]{\ensuremath{\bm{{#1}}}}
\newcommand{\ket}[1]{\ensuremath{\left|{#1}\right\rangle}}
\newcommand{\bra}[1]{\ensuremath{\left\langle{#1}\right |}}

\newcommand{\beq}{\begin{equation}}
\newcommand{\eeq}{\end{equation}}
\newcommand{\bse}{\begin{subequations}}
	\newcommand{\ese}{\end{subequations}}\newcommand{\bea}{\begin{eqnarray}}
\newcommand{\eea}{\end{eqnarray}}
\newcommand{\bit}{\begin{itemize}}
	\newcommand{\eit}{\end{itemize}}
\newcommand{\bpmatrix}{\begin{pmatrix}}
	\newcommand{\epmatrix}{\end{pmatrix}}

\newcommand{\be}{\begin{equation}}
\newcommand{\ee}{\end{equation}}
\newcommand{\ben}{\begin{eqnarray}}
\newcommand{\een}{\end{eqnarray}}

\begin{document}

\title{Behavior of quantum correlations under
nondissipative decoherence by means of the correlation matrix}

\author{D. G. Bussandri$^{1,2}$, T. M. Os\'an$^{1,3}$, A. P. Majtey$^{1,3}$, P. W. Lamberti$^{1,2}$}
\affiliation{$^1$Facultad de Matem\'atica, Astronom\'{\i}a, F\'{\i}sica y Computaci\'on, Universidad Nacional de C\'ordoba, Av. Medina Allende s/n, Ciudad Universitaria, X5000HUA C\'ordoba, Argentina}
\affiliation{$^2$Consejo Nacional de Investigaciones Cient\'{i}ficas y T\'ecnicas de la Rep\'ublica Argentina, Av. Rivadavia 1917, C1033AAJ, CABA, Argentina}
\affiliation{$^3$ Instituto de F\'isica Enrique Gaviola, Consejo Nacional de Investigaciones Cient\'{i}ficas y T\'ecnicas de la Rep\'ublica Argentina, Av. Medina Allende s/n, X5000HUA, C\'rdoba, Argentina}

\begin{abstract}
In this paper we use the Fano representation of two-qubit states from which we can identify a correlation matrix containing the information about the classical and quantum correlations present in the bipartite quantum state. To illustrate the use of this matrix, we analyze the behavior of the correlations under non-dissipative decoherence in two-qubit states with maximally mixed marginals. From the behavior of the elements of the correlation matrix before and after making measurements on one of the subsystems, we identify the classical and quantum correlations present in the Bell-diagonal states. In addition, we use the correlation matrix to study the phenomenon known as freezing of quantum discord. We find that under some initial conditions where freezing of quantum discord takes place, quantum correlation instead may remain not constant. In order to further explore into these results we also compute a non-commutativity measure of quantum correlations to analyze the behavior of quantum correlations under non-dissipative decoherence. We conclude from our study that freezing of quantum discord may not always be identified as equivalent to the freezing of the actual quantum correlations.
\end{abstract} 
\maketitle

\section{Introduction\label{sec:intro}}

In quantum information processing and quantum computing a central issue is to improve our capability of identifying which features in the quantum realm are responsible for the speed up of quantum algorithms over their classical counterparts. For a long time the prime suspect was the entanglement. However, it has been proven both theoretically and experimentally that there exists some separable mixed states, having negligible entanglement, which provide computational speedup in some quantum computation models compared to classical procedures [13,24]. Several results indicate that the increase in the efficiency is due to correlations of a quantum nature different from entanglement \cite{Knill98}-\cite{Lanyon08}.

Quantum discord (QD) is a widely accepted measure of quantum correlations, beyond just entanglement, and it is useful in many ways to indicate a divergence from classicality. However, even though there is strong evidence that states with non-zero discord play a central role in mixed state protocols, in the context of quantum state algorithms there is still interest in understanding the elusive source for the quantum speed up. Thus, besides QD, several measures of quantum correlations have been proposed \cite{ABC16}. 
Of special relevance for this work is the non-commutativity measure of quantum correlations (NCMQC) introduced in \cite{Guo16} and \cite{Majtey17}.

On the other hand, it is well-known that quantum correlations are usually destroyed under the effects of decoherence , i.e., uncontrolled interactions between the system and its environment. As a consequence, the system becomes less efficient for the realization of a number of quantum information tasks. Thus, in order to faithfully perform quantum information protocols it is of the essence to know the time scales along with the involved quantum resources can be securely preserved and manipulated.

Recent studies of the dynamics of general quantum correlations in open quantum systems under Markovian or non-Markovian evolutions indicate that QD is typically more robust than entanglement and does not suffer from sudden death issues \cite{Modi2012,Czele2011,LoFranco2013,Maziero2009,Ferraro2010}. In particular, a peculiar  phenomenon known as \textit{discord freezing} can occur for two-qubit states undergoing nondissipative decoherence. Indeed, under Markovian conditions and for certain initial conditons, QD may remain constant or frozen for a time interval \cite{Mazzola2010}. Moreover, when a non-Markovian dynamics is considered, a forever frozen discord \cite{Haikka2013} or multiple intervals of recurring frozen discord \cite{Mazzola2011,Mannone2013,LoFranco2012} may take place. Even though necessary and sufficient conditions for the freezing have been investigated \cite{You2012}, this phenomenon continues to be not completely understood.  Besides, it is natural to question whether the freezing phenomenon is a consequence of a mathematical artifact originated from the particular definition of QD or it reflects the actual freezing of quantum correlations present in the physical system. The aim of this work is precisely to gain more insights in order to answer this question.

This paper is organized as follows. In Sect.~\ref{sec:theory} we outline the theoretical framework for our work, including the definition of quantum discord, the Fano form and the correlation matrix for two-qubit states, the properties of Bell-diagonal states, and also the definition of a non-commutativity measure of quantum correlations. In Sect.~\ref{sec:Results}, we present our main results. We obtain the correlation matrix after a measurement has been performed on one of the subsystems. By using the correlation matrix we determine the character of the correlations present in a two-qubit Bell-diagonal state. By considering a dynamical scenario, corresponding to a non-dissipative decoherence process, we discuss the freezing phenomenon of quantum discord analyzing the behavior of the correlations given by the correlation matrix, the QD measure, and according to the non-commutativity measure of quantum correlations. Finally, some conclusions are addressed in Sect.~\ref{sec:conclusions}.

\section{Theoretical framework\label{sec:theory}}

\subsection{Quantum discord}

A widely accepted information-theoretic measure of the total correlations contained in a bipartite quantum state $\rho$ is the (von Neumann) Quantum Mutual Information $\mathcal{I}(\rho)$ defined as:
\begin{align}
\mathcal{I}(\rho) \doteq S(\rho_A) + S(\rho_B) -
S(\rho).
\label{eq:QMI}
\end{align}
In eq. \eqref{eq:QMI}, $\rho$ stands for a general bipartite quantum state, 
$\rho_A=\textrm{Tr}_B\left[\rho \right]$, $\rho_B=\textrm{Tr}_A\left[\rho \right]$ represent the corresponding reduced (marginal) states and $S(\rho)$ represents the von Neumann entropy given by
\begin{align}
S(\rho) \doteq -\mbox{Tr} \left[ \rho \log_2 \rho\right] .
\label{def8}
\end{align}
It is worth mentioning that $\mathcal{I}(\rho)$ describes the correlations between the whole subsystems rather than a  correlation between just two observables.\par
Classical correlations present in a quantum state $\rho$ of a bipartite quantum system can be quantified by means of the measure $\mathcal{J}_S(\rho)$ defined as ~\cite{OZ02,HV01}
\begin{align}
\mathcal{J}_S\left(\rho\right)\doteq S(\rho_B)-\min_{\mathcal{M}} \sum_j \ p'_j \ S (\rho_{B|j}^\mathcal{M} ), \label{eq:classcorrVedral}
\end{align}
\noindent with $\mathcal{M}=\left\{M_j\right\}_{j=1}^{m}$ ($m\in\mathbb {N} $) being a von Neumann measurement on subsystem $A$ (i.e., a complete set of rank-1 orthonormal projective measurements on $\mathcal{H}_A$), and 
\begin{align}
\rho_{B|j} ^\mathcal{M}&=\textrm{Tr}_A\left[( M_j \otimes \mathbb{I}) \rho \right]/p'_j \label{rhoBJ} \\
p'_j&=\textrm{Tr}\left[( M_j \otimes \mathbb{I}) \rho \right], \label{PprimaJ}
\end{align}
\noindent being the resulting state of the subsystem $B$ after obtaining the result $M_j$ when $\mathcal{M}$ is measured on subsystem $A$ and $p'_j$ being its corresponding probability. States given by Eq. \eqref{rhoBJ} are commonly referred to as \textit{conditional states}.\par
The difference between total correlations given by  $\mathcal{I}(\rho)$ [cf. Eq. \eqref{eq:QMI}] and classical correlations as measured by $\mathcal{J}_S(\rho)$ [cf. Eq. \eqref{eq:classcorrVedral}]  provides the measure of quantum correlations known as Quantum Discord which can be written as ~\cite{OZ02,HV01},
\begin{align}
\mathcal{D} (\rho) \doteq S(\rho_A) - S(\rho) + \min_{\mathcal{M}} \sum_j \ p'_j \ S (\rho_{B|j}^\mathcal{M} ).
\label{eq:QDdef}
\end{align}
Besides, after a measurement $\mathcal{M}$ is performed on party $A$, the state of the composite system $A+B$ (without observing) can be written as
\begin{align}\label{rhoM}
	\rho^\mathcal{M}=\sum_j (M_j\otimes\mathbb{I}) \rho (M_j\otimes\mathbb{I}). 
\end{align}
Thus, bearing in mind equation \eqref{rhoM}, it can be easily verified that Quantum Discord can also be written as
\begin{align}
\mathcal{D} (\rho) \doteq \mathcal{I}(\rho)  - \max_{\mathcal{M}}\mathcal{I}(\rho^\mathcal{M}). 
\label{eq:QDdef2}
\end{align}
It is worth pointing out that, as the measure $\mathcal{J}_S\left(\rho\right)$ is not symmetric under the exchange of subsystems $A$ and $B$, there exists a \textit{directionality} over $\mathcal{J}_S\left(\rho\right)$ and in consequence over the quantity $\mathcal{D}(\rho)$.

\subsection{Correlation matrix, classical and quantum correlations\label{sec:Tmatrix}}

A general two qubits state $\rho$ may always be written, up to local unitary transformations, in the Fano form \cite{Fano1983,Hioe1981,Schlienz1995} as follows
\begin{align}
\rho = \rho_A \otimes \rho_B + \frac{1}{4}\sum_{ij} T_{ij}\, \sigma^A_i\otimes \sigma^B_j.\label{Fanoform}
\end{align} 
Here $\rho_{A,B}=\textrm{Tr}_{B,A}[\rho]$; $\{\sigma^A_i\}$, $\{\sigma^B_i\}$ denotes the Pauli matrices acting on the Hilbert spaces $A$, $B$ respectively; and the elements
\begin{align}
T_{ij}\doteq\left\langle\sigma^A_i\otimes\sigma^B_j\right\rangle_\rho-\left<\sigma^A_i\otimes\mathbb{I}\right>_\rho\left<\mathbb{I}\otimes\sigma^B_j\right>_\rho,
\end{align}
\noindent define the \textit{correlation} matrix $\mathbb{T}$, being $\left<O\right>_\rho \doteq \textrm{Tr}[O\rho]$. \par

On one hand, by analogy with the concept of correlation functions for describing correlation effects in many-body physics \cite{Ma1985} and taking into account that correlation functions are directly related to observables \cite{Ma1985}, it can be verified that the information related to both, classical and quantum correlations present in the composite quantum system, is in fact contained inside the elements $T_{ij}$ of the correlation matrix $\mathbb{T}$. This matrix was used to investigate, for example, the dynamics of open quantum systems in the presence of initial correlations \cite{Peter2001}, and to study correlations in the quantum state of a composite system \cite{Huang2008,Dong2010}.

\subsection{Two-qubit states with maximally mixed marginals}

Bell--diagonal (BD) states are two-qubit states with maximally mixed marginals which can be written as
\begin{align}\label{BD states}
\rho^{BD}=\frac{1}{4}\left( \mathbb{I}_2 \otimes \mathbb{I}_2 + \sum_{i=1}^3 c_{i} \sigma^A_i \otimes \sigma^B_i \right),
\end{align} 
with $\mathbb{I}_2$ the identity matrix of dimension 2. 

Any two-qubit state satisfying $\langle\sigma_j^A\rangle=0=\langle\sigma_j^B\rangle$, i.e., having maximally mixed marginal density operators $\rho_A=\mathbb{I}_2/2=\rho_B$, can be brought into a Bell-diagonal form by using local unitary operations on the two qubits to diagonalize the correlation matrix $\langle\sigma_j^A\otimes\sigma_k^B\rangle$. Since quantum and classical correlations are both invariant under local unitary transformations, for our purpose it will be sufficient to consider the set of BD states.

The eigenvalues of a BD state are given by
\begin{align}
\lambda_0&=\frac{1}{4}(1-c_1-c_2-c_3),\\
\lambda_1&=\frac{1}{4}(1-c_1+c_2+c_3),\\
\lambda_2&=\frac{1}{4}(1+c_1-c_2+c_3),\\
\lambda_3&=\frac{1}{4}(1+c_1+c_2-c_3),
\end{align}
\noindent where the coefficients $\{c_j\}$ are such that $0\leq\lambda_i\leq 1$, $i=0,\ldots, 3$.  

BD states are a three-parameter set which includes the subsets of separable and classical states~\cite{Horodeckis2009}. They can be specified by the 3-tuple $(c_1,c_2,c_3)$. 
Two-qubit states with maximally mixed marginals also includes Werner ($|c_1|= |c_2|= |c_3|=c$) and Bell states ($|c_i|=1$, $|c_j|=0$, $|c_k|=0$, with $(i,j,k)$ any permutation of $(1,2,3)$). Thus, the state represented by Eq. \eqref{BD states} encompasses a wide set of quantum states.\par

\subsection{Non-commutativity measure of quantum correlations}\label{sec:NCMQC}

In \cite{Guo16} a non-commutativity measure of quantum correlations (NCMQC) was introduced as another tool for studying the behavior of quantum correlations in bipartite quantum systems.\par
Any state $\rho$ of a bipartite system $A+B$ can always be expressed as
\beq\label{rho}
\rho=\sum_{i,j}  A_{ij}\otimes |i_B\rangle\langle j_B|,
\eeq
\noindent where $\{|i_B\rangle\}$ stands for an orthonormal basis of $\mathcal{H}_B$, and 
\beq \label{As}
A_{ij}\doteq \mathrm{Tr}_B[(\mathbb{I}_A\otimes|j_B\rangle\langle i_B|)\rho].
\eeq 
By considering this representation of the states, Guo \cite{Guo16} introduced the following measure of quantum correlations: 
\beq
D_{A}(\rho) \doteq\sum_{\Omega}||[A_{ij},A_{kl}]||_2,\label{Dtraza}
\eeq
where $||\cdot||_2$ is the Hilbert-Schmidt norm, $||A||_2=\sqrt{\mathrm{Tr}(A^{\dagger}A)}$, and $\Omega$ the set of all the possible pairs (regardless of the order). 

According to \cite{Majtey17}, the measure $D_A(\rho)$ depends upon the representation basis $\{\ket{i_B}\}$ of the state $\rho$ \eqref{rho}. Then, it fails in satisfying all the criteria in order to be a physically well-behaved measure of quantum correlations. With the aim of overcoming this drawback, the following improved measure of quantum correlations has been proposed \cite{Majtey17}:
\begin{align}
\label{dnos}
d_{A}(\rho) \doteq \min_{\mathcal{R}}D_{A}(\rho),
\end{align}
where the minimum is taken over all possible representations of the state $\rho$.

\section{Results}\label{sec:Results}
In this section we analyze the (quantum or classical) character of the correlations present in a bipartite state $\rho$ by means of the Fano representation {\eqref{Fanoform}} and the correlation matrix $\mathbb{T}$. We shall focus on two-qubit BD states.
\subsection{Correlation matrix as a tool to identify classical and quantum correlations. \label{sec:correMatr}}

The computation of the QD involves an optimization of the \textit{classical correlations} $\mathcal{J}_S$ [cf. eq. \eqref{eq:classcorrVedral}] over all possible von Neumann measurements. Let us introduce  local measurements for party $A$,
\begin{equation}
\{E_j=\ket{j}\bra{j} \ / \ j\in\{0,1\}\},
\end{equation}
that is, $\{E_j\}$ is a PVM (Projection-Valued Measure) over the subsystem $A$ given in the computational basis $\{\ket{j}\}$. Any other projective measurement will be given by a unitary transformation:
\begin{equation}\label{param1}
\{M_j=V\ket{j}\bra{j}V^\dagger \ / \ j\in\{0,1\}\},
\end{equation}
with $V\in U(2)$. A useful parametrization of this unitary operators, up to a constant phase, is
\begin{align}\label{param2}
V=\vect{s}\cdot(\mathbb{I}_2,i \vect{\sigma}),
\end{align}
with $\vect{s}\in \Gamma$, and $\Gamma =\{\vect{s}\in\mathbb{R}^4 \ / \ s_0^2+s_1^2+s_2^2+s_3^2=1\}$. 

Once the measurement is parametrized by
the vector $\vect{s}$, and considering Bell diagonal states \eqref{BD states}, the conditional states of the subsystem $B$ [cf. Eq. \eqref{rhoBJ}] are given by \cite{Luo08b}

\begin{align}
\rho^{BD}_{B|0}(\vect{s})&=\frac{1}{2}\left(\mathbb{I}_2 + \sum_{i=1}^3 c_i z_i(\vect{s}) \sigma^B_i\right),\label{eq:rhob0}\\
\rho^{BD}_{B|1}(\vect{s})&=\frac{1}{2}\left(\mathbb{I}_2 - \sum_{i=1}^3 c_i z_i(\vect{s}) \sigma^B_i\right),\label{eq:rhob1}
\end{align}
In Eqs. \eqref{eq:rhob0} and \eqref{eq:rhob1} we defined
\begin{align}
z_1(\vect{s})&=2(-s_0s_2+s_1s_3)\label{eq:z1},\\
z_2(\vect{s})&=2(s_0s_1+s_2s_3)\label{eq:z2},\\
z_3(\vect{s})&=s_0^2+s_3^2-s_1^2-s_2^2,\label{eq:z3}
\end{align}

\noindent and the conditional probabilities are $p_{0}(\vect{s})\!=\!p_{1}(\vect{s})\!=\!\frac{1}{2}$ for all $\vect{s}\in \Gamma$. \par
By using \eqref{eq:z1}, \eqref{eq:z2}, and \eqref{eq:z3} the measure $\mathcal{J}_S$ [cf. Eq. \eqref{eq:classcorrVedral}] of classical correlations can be evaluated. It turns out that $\mathcal{J}_S$ is a non-decreasing function of the parameter $\theta(\vect{s}):=\sqrt{ |c_1z_1(\vect{s})|^2+|c_2z_2(\vect{s})|^2+|c_3z_3(\vect{s})|^2}$. Therefore, the \textit{optimal measurement} is defined by the vector $\vect{s}$ such that $\theta(\vect{s})$ is maximum. \par
If we set $c=\max\{|c_1|,|c_2|,|c_3|\}$ it can be verified that $\theta(\vect{s}) \leq c$. Thus, the optimal measurement is given by the vector $\vect{s}_M$ satisfying $\theta(\vect{s}_M)=c$. More specifically, we have the following cases,\par

\begin{enumerate}
	\item If $c=|c_1|$ $\Rightarrow$ $|z_1(\vect{s}_M)|=1$, $z_2(\vect{s}_M)=z_3(\vect{s}_M)=0$;
	\item If $c=|c_2|$ $\Rightarrow$ $|z_2(\vect{s}_M)|=1$, $z_1(\vect{s}_M)=z_3(\vect{s}_M)=0$;
	\item If $c=|c_3|$ $\Rightarrow$ $|z_3(\vect{s}_M)|=1$, $z_2(\vect{s}_M)=z_1(\vect{s}_M)=0$.
\end{enumerate}

The correlation matrix for an arbitrary BD state [cf. \eqref{BD states}] takes the form,
\begin{align}
	\mathbb{T}=\begin{bmatrix}
	c_1 & 0 & 0 \\
	0 & c_2 & 0 \\
	0 & 0 & c_3 
	\end{bmatrix},
\end{align} 
revealing the presence of correlations between the Pauli spin observables $\sigma^A_i$ and $\sigma^B_i$ of each subsystem, $i\in \{1,2,3\}$. \par
After some algebra, when a measurement parametrized by equations (\ref{param1}) and (\ref{param2}) is performed on subsystem $A$, it is straightforward to verify that the correlation matrix associated with the state after the measurement can be written as,
\begin{align}
\mathbb{T}(\vect{s})=\begin{bmatrix}
c_1z_1(\vect{s})^2 & c_2z_1(\vect{s})z_2(\vect{s}) & c_3z_1(\vect{s})z_3(\vect{s}) \\
c_1z_1(\vect{s})z_2(\vect{s}) & c_2z_2(\vect{s})^2 & c_3z_2(\vect{s})z_3(\vect{s}) \\
c_1z_1(\vect{s})z_3(\vect{s}) & c_2z_2(\vect{s})z_3(\vect{s}) & c_3z_3(\vect{s})^2 
\end{bmatrix}
\end{align} 
If we choose $\vect{s}$ maximizing $\mathcal{J}_S$, i.e,. $\vect{s}=\vect{s}_M$, the correlation matrix becomes diagonal with only one non--vanishing element given by $c=\max\{|c_1|,|c_2|,|c_3|\}$. For example, 
\begin{itemize}
	\item if $c=|c_1|$,
	\begin{align}\label{matrixc1}
	\mathbb{T}(\vect{s}_M)=\begin{bmatrix}
	c_1 & 0 & 0 \\
	0 & 0 & 0 \\
	0 & 0 & 0 
	\end{bmatrix},
	\end{align} 
	\item if $c=|c_2|$,
	\begin{align}\label{matrixc2}
	\mathbb{T}(\vect{s}_M)=\begin{bmatrix}
	0 & 0 & 0 \\
	0 & c_2 & 0 \\
	0 & 0 & 0 
	\end{bmatrix},
	\end{align} 
	\item if $c=|c_3|$,
	\begin{align}
	\mathbb{T}(\vect{s}_M)=\begin{bmatrix}
	0 & 0 & 0 \\
	0 & 0 & 0 \\
	0 & 0 & c_3 
	\end{bmatrix}.
	\end{align} 
\end{itemize}

Taking into account that after a measurement the state can only exhibit classical correlations \cite{Majtey17,HV01,ABC16}, the elements in the correlation matrix which remain invariant after the (optimal) measurement can be associated with this kind of correlations \cite{HV01,OZ02,Luo08b}. Thus, BD states exhibit only one classical correlation in the direction determined by  $c=\max\{|c_1|,|c_2|,|c_3|\}$. On the other hand, those elements of $\mathbb{T}$ suppressed by the measurement can be identified with quantum correlations. 

\subsection{Behavior of correlations under nondissipative decoherence\label{sec:decores1}}
Now, we turn to the study of a dynamical scenario where we shall consider two non-interacting qubits $A$ and $B$ under the influence of local and identical non-dissipative decoherence channels. In this case, the evolution of a two-qubit state $\rho$ can be written by means of the Kraus operators formalism, e.g., 	
\begin{align}
	\Lambda[\rho]=\sum_{i,j=1}^4(E^A_i\otimes E^B_j) \rho (E_i^{A\dagger} \otimes E_j^{B\dagger}),
\end{align}
where the Kraus operators are
\begin{align}
	&E_{k}^m=\sqrt{\frac{1-\exp(-\gamma t)}{2}}\sigma^m_k, \\
	&E_{4}^m=\sqrt{\frac{1+\exp(-\gamma t)}{2}}\mathbb{I}_2, \\
	&E_{i,j \not= k}^m=0,
\end{align}

\noindent and $m=A,B$ states for the qubit $A$ or $B$, $k\in\{1,2,3\}$ is in correspondence with $\{$\textit{bit flip}, \textit{bit-phase flip}, \textit{phase flip}$\}$ channels, and $\gamma\in\mathbb{R}_{\geq0}$ is the decoherence rate. A particular choice of $k$ defines the direction $x$, $y$, $z$ of the noise in the Bloch sphere and establishes the decoherence process.

 If the $A+B$ system is initially  in a BD state its structure will remain unchanged for all  $t$ \cite{Modi2012,Czele2011,LoFranco2013,Maziero2009,Ferraro2010}. In this scenario, the coefficients $c_i$ are functions of $t$ and are given by
\begin{align}
	&c_k(t)=c_k(0), \\
	&c_{i,j \not= k}(t)=c_{i,j}(0)e^{-2\gamma t}.
\end{align}
The freezing phenomenon of QD may occur if  certain particular initial conditions are satisfied, as for example:
\begin{align}
	c_i(0)&=\pm 1, \\
	c_j(0)&=\mp c_k(0),
\end{align}  
with $\left|c_k(0)\right|\equiv c_0$ and $k\in\{1,2,3\}$ denoting the corresponding channels.

The evolution of the system from the above initial conditions gives rise to a peculiar dynamics. In particular, some measures of quantum correlations \cite{Cianciaruso2015}, remain constant for all $t\in [0,t^*]$ where $t^*=-\frac{1}{2\gamma}\log c_0$. However, for $t>t^*$ they start to decay with $t$.\par

\subsubsection{Correlation matrix}
Here we analyze the dynamics of quantum	and classical correlations under a non-dissipative decoherence process by using	the results of Sect.~\ref{sec:correMatr}.

In the case of the phase flip channel ($k=3$), the correlation matrix for a BD state can be written as 
\begin{align}
\mathbb{T}=\begin{bmatrix}
c_{10}e^{-2\gamma t} & 0 & 0 \\
0 & c_{20}e^{-2\gamma t} & 0 \\
0 & 0 & c_{30} 
\end{bmatrix},
\end{align}
with $c_j(0)=c_{j0}$, $j\in\{1,2,3\}$. After performing the optimal measurement, the structure of the correlation matrix will be determined by $$c(t)=\max\{|c_1(t)|,|c_2(t)|,|c_3(t)|\}$$ which in turn will depend upon the initial conditions. \par

In order to illustrate the use of the correlation matrix for the analysis of the correlations present in the system in what follows we shall consider two examples corresponding to two different sets of initial conditions.\par
\begin{figure}
	\centering
	\includegraphics[width=0.75\textwidth]{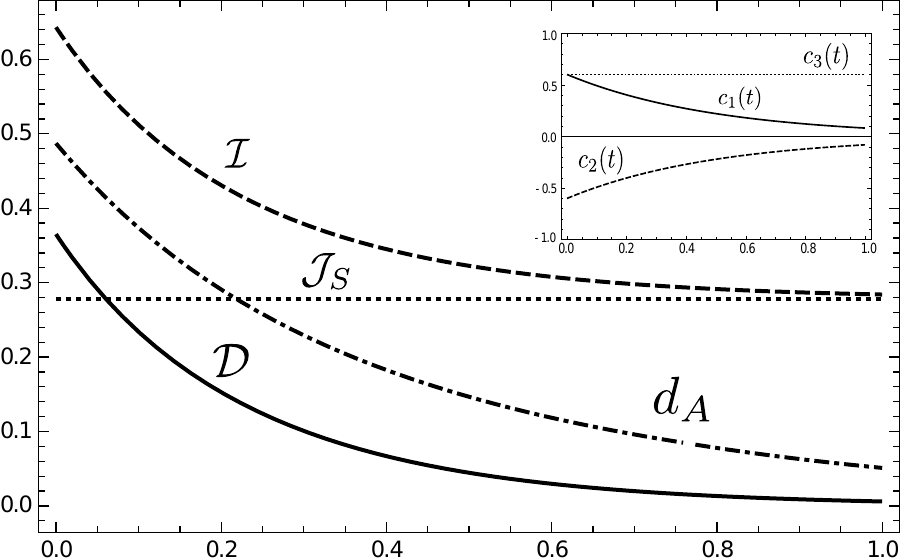}
	\caption{Dynamics of $\mathcal{D}$ (solid line), $\mathcal{I}$ (dashed line), $\mathcal{J}_S$ (dotted line) and $d_A$ (dash-dotted line) as a function of $t$ ($\gamma=1$) for $c_1(0)=-c_2(0)=c_3(0)=0.6$ and $k=3$. All shown quantities are dimensionless.}
	\label{Fig1}
\end{figure}

\vspace{0.5cm}
\noindent \textit{Example 1:} Let us consider the simple case where we set $c_{10}=c_0$, $c_{20}=-c_0$ and $c_{30}=c_0$. As $c_1(t)$ and $c_2(t)$ both decay with time, it is clear that $c(t)=c_0$. In this case we have,
\begin{align}
\mathbb{T}=\begin{bmatrix}
c_0e^{-2\gamma t} & 0 & 0 \\
0 & -c_0e^{-2\gamma t} & 0 \\
0 & 0 & c_0 
\end{bmatrix},
\end{align}
and after performing the optimal measurement the correlation matrix takes the form,
\begin{align}
\mathbb{T}(\vect{s}_M)=\begin{bmatrix}
0 & 0 & 0 \\
0 & 0 & 0 \\
0 & 0 & c_0 
\end{bmatrix},
\end{align}
for all $t\in[0,\infty)$. Therefore, after the measurement, the invariant element turns out to be $T_{33}=c_0$. Thus this element is associated with a classical correlation. In contrast, as the elements $c_1(t)$, $c_2(t)$ are suppressed under measurement, they are associated with quantum correlations. Here we consider the optimal measurement correspondig to the computation of the classical correlations $\mathcal{J}_S$. Following our analysis, quantum discord $\mathcal{D}$ should also decay and the classical correlations should remain constant. This matches perfectly  with the behaviour of correlations shown in Fig. \ref{Fig1}.\par

\begin{figure}
	\centering
	\includegraphics[width=0.75\textwidth]{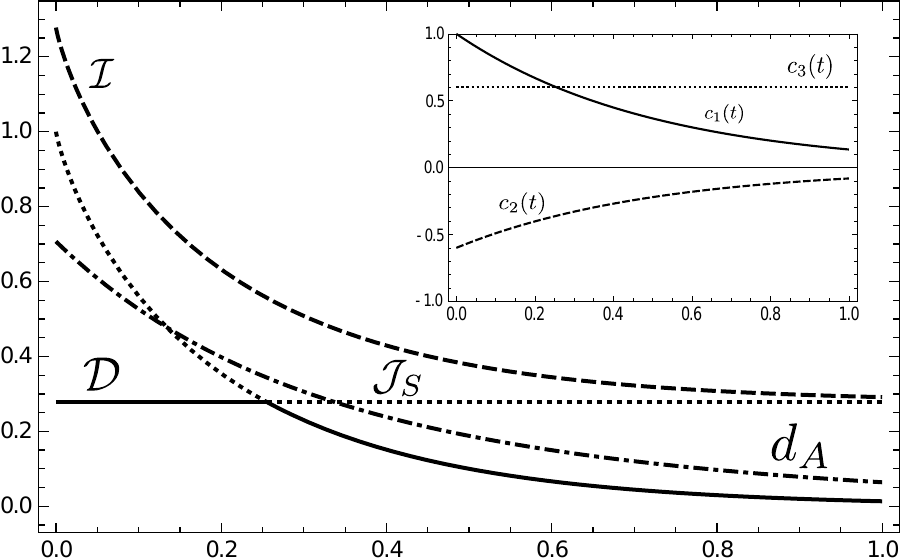}
	\caption{Dynamics of $\mathcal{D}$ (solid line), $\mathcal{I}$ (dashed line), $\mathcal{J}_S$ (dotted line) and $d_A$ (dash-dotted line) as a function of $t$ ($\gamma=1$) for $c_1(0)=1$, $c_3(0)=-c_2(0)=0.6$ and $k=3$. All shown quantities are dimensionless.}
	\label{Fig2}
\end{figure}

\vspace{0.5cm}
\noindent  \textit{Example 2:} Now, let us consider the freezing phenomenon of quantum discord. In this case, we set the initial conditions as follows: $c_{10}=1$, $c_{20}=-c_0$, and $c_{30}=c_0$. The behavior of the (total, classical, and quantum) correlations measures and the matrix elements are plotted in Fig. \ref{Fig2}. 
 In this case there exists a clear change in the correlation matrix $\mathbb{T}(\vect{s}_M)$ at $t^*$ and a sharp transition of quantum and classical correlations is also observed. The correlation matrix takes the form, 
\begin{align}
\label{matrixTfreezing}
\mathbb{T}=\begin{bmatrix}
e^{-2\gamma t} & 0 & 0 \\
0 & -c_0e^{-2\gamma t} & 0 \\
0 & 0 & c_0 
\end{bmatrix},
\end{align}
and after performing  the optimal measurement, for $t\in (t^*,\infty)$ we have $c(t)=|c_3|=|c_0|$ leading to the following matrix, 

\begin{align}
\mathbb{T}(\vect{s}_M)=\begin{bmatrix}
0 & 0 & 0 \\
0 & 0 & 0 \\
0 & 0 & c_0 
\end{bmatrix}.
\end{align}
Thus, $T_{33}=c_0$ is associated with a classical correlation whereas the remaining elements $T_{11}=e^{-2\gamma t}$, $T_{22}=-c_0e^{-2\gamma t}$ are associated with quantum correlations. This analysis is in agreement with Fig. \ref{Fig2} where it can be seen that while the measure of classical correlations remains constant, quantum discord decays with $t$. However, for $t\in [0,t^*)$ we have $c(t)=|c_1(t)|$. Therefore, the correlation matrix takes form,
\begin{align}
\label{matrixTfreezingCC2}
\mathbb{T}(\vect{S}_M)=
\begin{bmatrix}
e^{-2\gamma t} & 0 & 0 \\
0 & 0 & 0 \\
0 & 0 & 0 
\end{bmatrix}.
\end{align}

\noindent If we compare Eqs. \eqref{matrixTfreezingCC2} and (\ref{matrixTfreezing}), we can see that now $T_{11}=e^{-2\gamma t}$ is associated with a classical correlation and $T_{22}=-c_0e^{-2\gamma t}$, $T_{33}=c_0$ are associated with quantum correlations. The measure of classical correlations decays in time in the same way as von Neumann total information does. Thus, quantum discord remains constant in this case exhibiting the freezing phenomenon. By analyzing the corresponding correlation matrix this behavior seems to be controversial because only one  of the elements associated to quantum correlations remains invariant with $t$ ($T_{33}$). Since quantum discord is also a function of the elements of $\mathbb{T}$, a question that naturally arises is whether QD truly reflects what happens with the \textit{actual} quantum correlations in this time interval.

\subsubsection{NCMQD for Bell diagonal states\label{sec:opt}}

In order to further explore into the results obtained in previous section, we will compute now the non-commutativity measure of quantum correlations introduced in Sect.~\ref{sec:NCMQC} [cf. Eq.\eqref{dnos}] for Bell diagonals states [cf. Eq.\eqref{BD states}] under the influence of local non-dissipative decoherence. We will consider the same two examples as in previous section.

Following \cite{Bussandri19} we have,

\begin{align}
&\left[A_{ij},A_{kl}\right]=\frac{i}{8}\Big[ c_1c_2\alpha^{(12)}_{ijkl}\sigma_3+c_1c_3\alpha^{(31)}_{ijkl}\sigma_2+c_2c_3\alpha^{(23)}_{ijkl}\sigma_1  \Big],
\end{align}
with $\alpha_{ijkl}^{(mn)}=\sigma_m^{ij} \sigma_n^{kl} -  \sigma_n^{ij} \sigma_m^{kl}$. Thus, after some algebra the HS norm of the commutators can be written as
\begin{align}\label{26}
\begin{Vmatrix}
\left[A_{ij},A_{kl}\right]
\end{Vmatrix}_2^2&=\frac{1}{2^5}\Big( 
\left|c_1c_2\right|^2\left|\alpha_{ijkl}^{(12)} \right|^2 +\left|c_1c_3\right|^2\left|\alpha_{ijkl}^{(31)} \right|^2 +\left|c_2c_3\right|^2\left|\alpha_{ijkl}^{(23)} \right|^2\Big),
\end{align}
where $\sigma_k^{ij}=\bra{i_B}\sigma_k \ket{j_B}$. 

The optimization procedure involved in NCMQC requires a suitable parametrization of the basis of $\mathcal{H}_B$. In this case, we have considered $\{\ket{i_B}\}=\{U\ket{i_B}_c\}$, with $\{\ket{i_B}_c\}$ being the computational basis and $U$ an unitary operator which can also be parametrized according to Eq. \eqref{param2}.

After some straightforward calculations, the resulting expression to be minimized turns out to be

\begin{align}
D_A(\rho)=&\frac{1}{\sqrt{2^3}} \sqrt{c_2^2c_3^2z_1(\vect{s})^2 + c_1^2c_3^2z_2(\vect{s})^2 + c_2^2c_1^2z_3(\vect{s})^2 }+\nonumber\\
+&\frac{1}{\sqrt{2}}\sqrt{c_2^2c_3^2\zeta_1(\vect{s}) + c_1^2c_3^2\zeta_2(\vect{s}) + c_2^2c_1^2\zeta_3(\vect{s}) }\label{eq:DANCMQD},
\end{align}
with $z_1(\vect{s})$, $z_2(\vect{s})$ and $z_3(\vect{s})$ defined according to Eqs. \eqref{eq:z1}, \eqref{eq:z2}, and \eqref{eq:z3} respectively and $\zeta_i(\vect{s})=1-z_i(\vect{s})^2, i=1,2,3$.

After some algebra (see appendix \ref{sec:optimNCMQD}) the optimized measure can be written as:
\begin{equation}
\begin{split}
d_A(\rho)=\frac{1}{\sqrt{2^3}}\min \{ \ &|c_1c_2|+2\sqrt{(c_2c_3)^2+(c_1c_3)^2} , \\ &|c_2c_3|+2\sqrt{(c_1c_2)^2+(c_1c_3)^2}, \\ &|c_1c_3|+2\sqrt{(c_1c_2)^2+(c_2c_3)^2} \ \}.
\label{eq:dArhooptim}
\end{split}
\end{equation}

We evaluate Eq. \eqref{eq:dArhooptim} for the initial conditions corresponding to both of the examples considered before. In the first case, we choose  $c_{10}=c_0$, $c_{20}=-c_0$ and $c_{30}=c_0$, and the freezing phenomenon of QD is absent. In the second case, we choose $c_{10}=1$, $c_{20}=-c_0$, and $c_{30}=c_0$, and the freezing phenomenon of QD takes place. The dynamics of the NCMQC for each set of the initial conditions is shown in Fig. \ref{Fig1} and \ref{Fig2}. The behavior exhibited by this measure seems to follow more reliably the behaviour shown by the correlations associated with the elements of the matrix $\mathbb{T}$ than the measure of QD.

\section{Concluding remarks\label{sec:conclusions}}

In this paper we studied the behavior of correlations under non-dissipative decoherence in two-qubit states with maximally mixed marginals by means of the Fano representation which allows us to identify a correlation matrix.  From the behavior of the elements of this correlation matrix before and after making measurements on one of the subsystems, we have been able to identify the classical and quantum correlations present in the bipartite states. In addition, we used the correlation matrix to study the phenomenon of freezing of quantum discord under non-dissipative decoherence. We found that under some initial conditions, where freezing of quantum discord takes place, the actual quantum correlations may not remain constant. In order to obtain further insights into these findings we also computed a non-commutativity measure of quantum correlations in the same dynamical scenario. We conclude from our study that freezing of quantum discord may not always be identified as equivalent to the freezing of the actual quantum correlations.\par
Naturally, our conclusions may be extended to other measures or quantifiers of quantum correlations eventually reflecting the same kind of freezing behavior. It seems that caution must be exercised regarding the interpretation of freezing of a certain measure of quantum correlations as equivalent to the freezing of the actual quantum correlations present in the physical system.
\begin{acknowledgements}
D.B., T.M.O, A.P.M., and P.W.L. acknowledge the Argentinian agency SeCyT-UNC and CONICET for financial support. D. B. has a fellowship from CONICET.
\end{acknowledgements}

\section*{Appendix: Optimization of the NCMQD}
\label{sec:optimNCMQD}

In order to minimize Eq. \eqref{eq:DANCMQD} let us consider the function $f(\vect{x})$:
\begin{align}
f(\vect{x})=&\sqrt{\sum_{i=1}^3 \alpha_i x_i^2} + 2\sqrt{\sum_{i=1}^3 \alpha_i (1-x_i^2)}, \\
g(\vect{x})=&\sum_i x_i^2=1\label{eq:constfx},
\end{align}
with $g(\vect{x})=1$ and $\vect{x}=(x_1,x_2,x_3)$. Up to a constant factor, $D_A(\rho)$ is a particular case of $f(\vect{x})$: $\alpha_1=(c_2c_3)^2$, $\alpha_2=(c_1c_3)^2$, $\alpha_3=(c_2c_1)^2$ and the variables $\vect{x}=(x_1,x_2,x_3)$ represent the quantities $\vect{z}=( z_1(\vect{s}) ,z_2(\vect{s}),z_3(\vect{s})  )$.

Following the method of Lagrange multipliers we have
\beq
\frac{\partial f}{\partial x_p}=\lambda\frac{\partial g}{\partial x_p},
\ee
where $p\in\{1,2,3\}$. Thus, we can write the following equations:
\begin{align}
\frac{\alpha_k x_k}{\sqrt{\theta}}-2\frac{\alpha_k x_k}{\sqrt{\alpha - \theta}}=2 \lambda x_k \label{eq:xk}, \\ 
\frac{\alpha_i x_i}{\sqrt{\theta}}-2\frac{\alpha_i x_i}{\sqrt{\alpha - \theta}}=2 \lambda x_i \label{eq:xi}, \\
\frac{\alpha_j x_j}{\sqrt{\theta}}-2\frac{\alpha_j x_j}{\sqrt{\alpha - \theta}}=2 \lambda x_j \label{eq:xj},
\end{align}
where $\theta=\sum_{p} \alpha_p x_p^2$, $\alpha=\sum_p \alpha_p $ and $i, j, k$ ($i\neq j \neq k$) are numbers belonging to the set $\{1,2,3\}$.

Without loss of generality, we may assume that $\alpha_k$, $\alpha_i$ and $\alpha_j$ are different from zero. In view of the constrain $g(x_1,x_2,x_3)=1$ [cf. Eq. \eqref{eq:constfx}], let us suppose that $x_k=0$, $x_i=0$ and $x_j=1$. Then Eqs. $\eqref{eq:xk}$ and $\eqref{eq:xi}$ are satisfied and Eq. $\eqref{eq:xj}$ can be fulfilled by taking
\beq
\lambda=\frac{\alpha_j}{2\sqrt{\theta}}-\frac{\alpha_j}{\sqrt{\alpha - \theta}}. \label{eq:xj2}
\ee
Thus, bearing in mind the permutation of $i,j,k$, we obtain three extremal points, i.e., $\vect{x}_e\in \{(0,0,1),(0,1,0),(1,0,0)\}$.

Let us take now $x_k=0$, and $x_i\not =0$, $x_j\not =0$. Then, we have the three quantities $x_i$, $x_j$ and $\lambda$ to be determined taking into account \eqref{eq:xi}, \eqref{eq:xj} and $x_i^2+x_j^2=1$. Accordingly,
\begin{align}
	\lambda=\frac{\alpha_i}{2}\left(\frac{1}{\sqrt{\theta}}-\frac{2}{\sqrt{\alpha-\theta}}\right).
\end{align}
Now, if $\alpha_i \not = \alpha_j$ from Eq. \eqref{eq:xj} we obtain
\begin{align}
	\frac{1}{\sqrt{\theta}}=\frac{2}{\sqrt{\alpha-\theta}}\label{eq:thetaeq}.
\end{align}
The equality in Eq. \eqref{eq:thetaeq} holds iff $\theta=\frac{1}{5}\alpha$. Therefore, the extremal points are given by
\begin{align}
	x_k&=0, \\
	x_1^2+x_j^2&=1, \\
	\alpha_ix_i^2+\alpha_jx_j^2&=\frac{1}{5}\sum_{p=1}^3\alpha_p.
\end{align}
On the contrary, if $\alpha_i = \alpha_j$, Eqs. \eqref{eq:xi} and \eqref{eq:xj} are trivially fulfilled. It can be verified that the general case, i.e., $x_k \not =0$, $x_i \not =0$ and $x_j \not =0$, can be solved following the previous calculations and gives the same extreme value $\theta=\frac{1}{5}\alpha$.

In summary, we have two types of extremal points. First $\vect{x}_e\in\{(0,0,1),(0,1,0),(1,0,0)\}$, secondly, $\vect{x}_e$ such that $\sum_p x_p^2=1$ and $\theta=\frac{1}{5}\alpha$. 

Let us see which of them is a minimum of the function $f(\vect{x})$. Consider the one-dimensional function $\hat{f}(\theta)=\sqrt{\theta}+2\sqrt{\alpha-\theta}$. It is easy to see that is a concave function of  $\theta$ with an extremal point in $\theta=\frac{1}{5}\alpha$. Therefore, this case corresponds to a local maximum. Thus, the minimum of the function should be in the boundary points:
\begin{align}
	\theta_{\min}=\min_{\vect{x}\in\mathcal{G}} \{ \theta \},\\
	\theta_{\max}=\max_{\vect{x}\in\mathcal{G}} \{\theta\},
\end{align}	
being $\mathcal{G}=\{\vect{x}\in \mathbb{R}^3 : g(\vect{x})=1 \}$. Following \cite{Luo08b}, as $\theta=\sum_p \alpha_p x_p^2\leq c\sum_p x^2_p=c_{\pm}$ we have,
\begin{align}
\theta_{\min}=c_{-}=\min\{\alpha_1,\alpha_2,\alpha_3\},\\
\theta_{\max}=c_{+}=\max \{ \alpha_1,\alpha_2,\alpha_3 \}.
\end{align}	
These $\theta$ values do coincide with our first type of extremal points $\vect{x}_e$. As a consequence, the minimum of the function $f(\vect{x})$ turns out to be:
\begin{align}
	f_{\min}=\min \{ \ &\sqrt{\alpha_1}+2\sqrt{\alpha_2+\alpha_3} \ , \ \sqrt{\alpha_2}+2\sqrt{\alpha_1+\alpha_3} \ , \nonumber \sqrt{\alpha_3}+2\sqrt{\alpha_2+\alpha_1} \ \}.
\end{align}
Finally, the optimized measure $d_A(\rho)$ can be written as:
\begin{equation}
\begin{split}
d_A(\rho)=\frac{1}{\sqrt{2^3}}\min \{ \ &|c_1c_2|+2\sqrt{(c_2c_3)^2+(c_1c_3)^2} , \\ &|c_2c_3|+2\sqrt{(c_1c_2)^2+(c_1c_3)^2}, \\ &|c_1c_3|+2\sqrt{(c_1c_2)^2+(c_2c_3)^2} \ \}.
\end{split}
\end{equation}
It is important to realize that the extremal points $\vect{z}_e=(z_1^e,z_2^e,z_3^e)\in\{ (0,0,1),(0,1,0),(1,0,0) \}$ can always be attained by a suitable choice of $\vect{s}$ \cite{Luo08b}.

{}

\end{document}